\documentclass[conference, compsoc]{IEEEtran}
\IEEEoverridecommandlockouts
\usepackage{subfigure}
\usepackage{caption}
\usepackage{hyperref}

\usepackage{amsmath}
\usepackage{booktabs}
\usepackage{graphicx}
\usepackage{multirow}
\usepackage{courier}
\usepackage{multirow}
\usepackage{acronym}
\usepackage{color}

\begin{document}

\title{On the Ethereum Blockchain Structure: a Complex Networks Theory Perspective\footnotemark}

\author{\IEEEauthorblockN{Stefano Ferretti, Gabriele D'Angelo}
\IEEEauthorblockA{Department of Computer Science and Engineering (DISI), University of Bologna, Italy\\
\{s.ferretti, g.dangelo\}@unibo.it}
}

\maketitle

\footnotetext{The publisher version of this paper is available at \url{https://doi.org/10.1002/cpe.5493}.
\textbf{{\color{red} This is the pre-peer reviewed version of the following
article: ``Stefano Ferretti, Gabriele D'Angelo. On the Ethereum Blockchain Structure: a Complex Networks Theory Perspective. To appear in Concurrency and Computation: Practice and Experience (Wiley)''.}}}

\begin{abstract}
In this paper, we analyze the Ethereum blockchain using the complex networks modeling framework. Accounts acting on the blockchain are represented as nodes, while the interactions among these accounts, recorded on the blockchain, are treated as links in the network. Using this representation, it is possible to derive interesting mathematical characteristics that improve the understanding of the actual interactions happening in the blockchain. Not only, by looking at the history of the blockchain, it is possible to verify if radical changes in the blockchain evolution happened.\\
\end{abstract}


\section{Introduction}\label{sec:intro}

The blockchain is arguably one of those technologies that, nowadays, are raising high expectations in terms of possible application domains. It is a global ledger that records transactions efficiently and permanently on a chain of blocks~\cite{8342866}. Each block contains a set of transactions created and dispatched in the system. Furthermore, each block contains a timestamp, a link to the previous block and it is identified by its hash value. All transactions are signed and hashed via cryptographic hash functions. This structure thus provides an unforgeable log containing the history of all the transactions ever made. Nodes participating to the blockchain are connected via a Peer-to-Peer (P2P) network. Each node maintains a replicated version of the entire transaction history.

Several variants of blockchains exist. While Bitcoin still remains the most famous one by the public, Ethereum probably represents one of the most interesting solution. This is due to the fact that Ethereum provides a vast range of use case applications enabled by smart contracts~\cite{Ethereum,8328741,8355340,cryblock2019,DiPascale:2017,Shafagh:2017:TBA:3140649.3140656,Neisse:2017:BAD:3098954.3098958,DBLP:conf/mobisys/0003GP18}.
Ethereum is often described with the term "world computer", since this platform enables running distributed applications (i.e.~smart contracts) in a distributed manner. It provides a way to create self-executing and self-enforcing contracts. Their execution is triggered via transactions. Once generated, nodes in the P2P system execute the related code. This causes a change of the state. All this is recorded in the blockchain. Thus, through the blockchain all nodes synchronize their replicated state globally, in a manner that is fully verifiable by any system participant. That is why the distributed code run on the blockchain is referred as a smart contract. Once deployed, it cannot be modified. Hence, parties, that agree on the use of this code, are aware that there is no possibility to breach the agreement. (They can, of course, decide to not use that contract anymore, if for some reason the contract becomes obsolete.)

In Ethereum, smart contracts are considered internal accounts, that can interact among themselves and with externally owned accounts, which are in fact users that employ the system. Both these kinds of accounts have their own balance, expressed in a distributed currency referred as Ether. The Ether is the fuel for operating in Ethereum. Every transaction in Ethereum is made possible through a payment made by the clients of the platform to the machines executing the requested operations. This enables several applications, ranging from the exchange of cryptocurrencies, to financial applications, storing and management of tokens and digital assets, notary systems, identity management, voting systems, up to those application that require the traceability of resources and assets~\cite{8257877,8355340,GaoXTPD0S18,SchonhalsHG18}. Several works find many application domains in healthcare, supply chain, Internet of Things, etc.~\cite{8252043,8257877,7749510,JavaidSAS18,SelimiKANS18,DBLP:journals/corr/abs-1812-02909,10.1093/jamia/ocx068}.

In this paper, we provide an analysis of the Ethereum blockchain. In particular, we employ the modeling techniques of the complex network theory~\cite{8328741,8356459,8486401}.  We represent the flow of transactions happened in the blockchain (or a subset of the blockchain) as a network, where nodes are the Ethereum accounts (i.e.~external accounts or smart contracts). In the Ethereum scenario, a transaction can represent some cryptocurrency transfer, the creation of a smart contract, or the invocation of a contract~\cite{8486401}. Each transaction recorded in the blockchain corresponds to the creation of a new link in the network. The rationale behind the analysis is that complex networks provide appropriate modeling to represent a blockchain as a complex system, together with powerful quantitative measures for capturing the essence of its complexity~\cite{gda-jpdc-2017,Ferretti20131631,10.1007/978-3-319-96661-8_45,DBLP:conf/webist/BaumannFL14,DBLP:conf/socialcom/ReidH11,DBLP:conf/mobisys/JavaroneW18}. 

Varying the number of blocks considered to extract the recorded transactions, we obtain different networks, of different size and complexity. This influences the structure of the network. The investigation, made in this work, leads to observations and insights. For instance, while a majority of nodes has a low degree (i.e.~just few amount of links), that demonstrates a poor level of interactions in the blockchain, we notice the presence of several hubs with higher degrees. This information is important to recognize which are the main contributors to the blockchain evolution. While these nodes are important ones, at the same time, they are exposed to a lower level of anonymity, i.e.~if we are dealing with an external owned account, it might be easier to discover the identity of the address corresponding to a specific node~\cite{DBLP:conf/socialcom/ReidH11}.

The remainder of this paper is organized as follows. Section~\ref{sec:background} describes the background and the state of the art about the blockchain technologies, Ethereum and complex network analysis. Section~\ref{sec:study} presents the approach used to model the Ethereum blockchain, and the interactions recorded in this distributed ledger, as a complex network. Section~\ref{sec:results} presents results obtained from the analysis of the different extracted networks. Finally, Section~\ref{sec:conc} provides some concluding remarks.

\section{Background}\label{sec:background}

\subsection{What is a Blockchain}

A blockchain is a distributed ledger that records transactions in blocks~\cite{Antonopoulos:2014:MBU:2695500,D'Angelo:2018}. Each block contains a set of transactions and it has a link to a previous block, thus creating a chain of chronologically ordered blocks. Transactions within a block are assumed to have happened at the same time. In the typical scenarios, transactions record an exchange of digital currencies, but in fact they can be employed to record any kind of event.

What makes the blockchain technology appealing is that the combination of P2P systems, cryptographic techniques, use of distributed consensus schemes and pseudonymity ensure that the set of confirmed transactions becomes public, traceable and tamper-resistant.  The latter property is obtained by linking subsequent blocks together using cryptographic hash functions so that the modification of transaction data in a block~$B_i$ would change the hash that is contained in the subsequent block~$B_{i+1}$, thus altering the content of block $B_{i+1}$ and so on. The blockchain is replicated across multiple nodes in a P2P fashion. Therefore, any attempt to alter the blockchain would create an easily detectable inconsistency
of all replicas.

The blockchain uses digital pseudonyms (addresses) -- usually, a hash of a public key -- to provide some level of anonymity. Therefore, everyone can trace the activities of an entity with a given pseudonym, but it is computationally expensive (although not impossible) to associate a pseudonym back to a specific entity or individual.

\subsection{Ethereum}

Ethereum is a specific blockchain-based software platform that enables the possibility of building and running smart contracts and the so called Distributed Applications (DApps)~\cite{8342866}. Such platform is also the basis for a related virtual currency, called Ether. For the definition of smart contracts, Ethereum provides a Turing complete programming language that allows creating programs and running them on the blockchain~\cite{8356459}.

Ethereum operates using accounts and their balances, that change via state transitions. The state denotes the current balances of all accounts, plus other possible extra data. The state is not stored on the blockchain directly, but it is encoded and maintained by accounts in a separate data structure organized as a Merkle Patricia tree. As in all permissionless blockchains, in order to provide anonymity, accounts are pseudonymous and are linked to one or more addresses~\cite{hjj2014}. There are two types of accounts: externally owned accounts and contracts accounts. Externally owned accounts are controlled by people. Thus, similarly to Bitcoin, each person has his own private key, which is used in order to make transactions in the Ethereum blockchain. Conversely, contract accounts are controlled by some smart contract code. In other words, such accounts are some sort of cyber-entities, having their own balance, that can be triggered through some transactions, coming from an external account (or some other contracts). Once triggered, the code specified in the contract is executed. This code can in turn generate some other transactions. The presence of these smart contracts allows developers to use Ethereum as a general purpose framework to create DApps.

The Ether is the cryptocurrency asset employed in the Ethereum blockchain. In some extent, the Ether is the fuel for operating the distributed applications over Ethereum. Using this cryptocurrency, it is possible to make payments to other accounts or to the machines executing some requested operation. Ether thus enables running DApps, enabling smart contracts, generating tokens during Initial Coin Offering (ICOs), i.e.~a type of funding using cryptocurrencies, and also for making standard P2P payments. That's why Ethereum is also referred as ``programmable money''.

\subsection{Complex Networks Analysis}

Complex networks theory allows to analyze a given real or synthetic system, and to extract several mathematical properties that describe it. It is quite usual to represent P2P and distributed systems~\cite{gda-jpdc-2017}, communication networks~\cite{gda-hpcs-16,Ferretti20131631}, social networks~\cite{DeMichele2018}, biological and very other diverse phenomena as complex networks. In order to describe a phenomenon as a network, entities are usually represented as network nodes, while interactions among these entities are links that connect these nodes. Depending on the symmetric or asymmetric nature of the interaction, these links may be undirected or directed, respectively.

In what follows, we briefly introduce the the main metrics, typically employed in complex network theory, that will be used to study the Ethereum blockchain.

\subsubsection{Number of nodes}
This measure is the total amount of nodes in the network. In our case, that is the total amount of different accounts which were involved in some transactions, in the considered snapshot of the Ethereum blockchain.

\subsubsection{Degree distributions}
The degree of a node $x$ is the amount of links that connect $x$ with other nodes in the network (included $x$ itself, when a loop is performed). The degree counts the number of addresses a given address had interactions with (i.e.~it was involved in one or more transactions).

Weights can be associated to links and exploited to measure the so called weighted degrees. In this case, a weight is assigned to each link, measuring the amount of transactions between two addresses in the considered time range. Thus, the weighted degree is the summation of the weights of links of a given node.

\subsubsection{Distance}
The average distance is the average shortest path length in a network, i.e.~the average number of steps along the shortest paths for all possible pairs of nodes. Distances among nodes are calculated using the standard breadth-first search algorithm, which finds the shortest distance from a single source node to every other node in the network.

This metrics should be considered together with the clustering coefficient. In fact, these two metrics allow determining if the network is a small world or not (as it will be described in the following of this section).

\subsubsection{Clustering coefficient}
The clustering coefficient is a measure assessing how much nodes in a graph tend to cluster together. It measures to what extent friends of a node are friends of one another too. When two connected nodes have a common neighbor, this triplet of nodes forms a triangle. The clustering coefficient is defined as 
$$C = \frac{3 \times \text{number of triangles in the network}}{\text{number of connected triplets of nodes}}$$ where a ``connected triple'' consists of a single node with links reaching a pair of other nodes; or, in other words, a connected triple is a set of three nodes connected by (at least) two links~\cite{Ferretti20131631}. A triangle of nodes forms three connected triplets, thus explaining the factor of three in the formula. In this context, a triangle of nodes means that an address $x$ had some transactions with other two, say $y$, $z$, and at the same time $y$ and $z$ had some transactions as well.

\subsubsection{Small worlds}
Small world networks are networks that are ``highly clustered, like regular lattices; yet, they have small characteristic path lengths, like random graphs''. In a small world, most nodes are not linked with each other, but most nodes can be reached from every other by a small number of hops. Indeed, in a small-world network the typical distance between two randomly chosen nodes grows proportionally to the logarithm of the number of nodes.

Given a network, it is possible to verify if it is a small world, by comparing it with a random graph of the same size. A random graph is a network with links randomly generated, based on a simple probabilistic model~\cite{Ferretti20131631}. Different models can be employed to generate a random graph. According to one of the simplest methods, a random graph can be constructed by creating a set of $n$ isolated nodes; then, we consider every possible pair of nodes $x$, $y$, and we add a link $(x,y)$ with probability $p$, independently of other links. Random graphs exhibit a small average distance among nodes (varying typically as the logarithm of the number of nodes, $\sim \ln(n)$) along with a small clustering coefficient $\sim\frac{\text{mun links}}{n^2}$.

In practice, one can assess whether a network has a small average distance as for a random graph, but a significantly higher clustering coefficient. In this case, the network is a small world. In particular, if one looks at the clustering coefficient ($cc$) together with the average distance ($L$) of the considered network, and the clustering coefficient ($cc_{RG}$) together with the average distance ($L_{RG}$) of the corresponding random graph, it is possible to measure the small-coefficient as
\begin{equation}\label{eq:sigma}
\sigma=\frac{cc/cc_{RG}}{L/L_{RG}},\end{equation}
concluding that the network can be classified as a small world when $\sigma$ is significantly higher than $1$.

\section{The Ethereum Blockchain as a Network}\label{sec:study}

It is possible apply the complex network machinery for the analysis of a blockchain. In this case, accounts that interact in the blockchain can be represented as network nodes, while their interactions can be seen as links. More specifically, the interactions represent transactions among different accounts. It is also possible to associate a weight to each link, that may further characterize the interaction. For instance, a counter may be associated to track the number of transactions made in the time interval of consideration. Alternatively, it might represent some other value, such as the currency being transferred between the two accounts.

\begin{figure*}
  \centering
  \includegraphics[width=13cm]{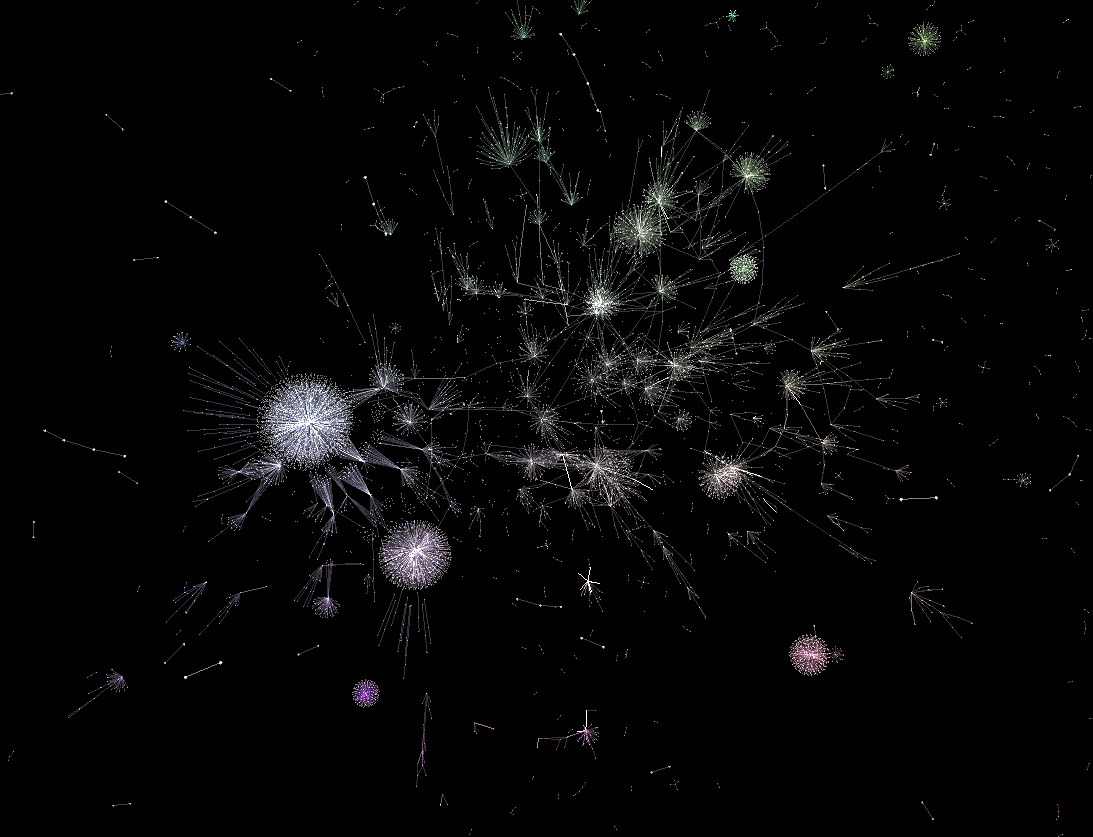}
  \caption{A screenshot of the appearing Ethereum interactions, as a network -- 1 hour log (approximately).}
  \label{fig:snap}
\end{figure*}

Figure~\ref{fig:snap} shows a screenshot of the 3D visualization of approximately 1-hour log of interactions (i.e.~240 blocks) in the Ethereum blockchain network. This figure shows some interesting indications about the network of transactions. In fact, there are several important nodes that are involved in many transactions, while the majority of nodes seem to have a single link entering/exiting from them. This is quite reasonable, since at the current status of this blockchain and its related cryptocurrency, it seems to be unlikely that a typical account participates in more that one transaction per hour. The analyzed network is composed of $28867$ nodes, corresponding to the same amount of accounts that have been active in the single hour being considered. The number of links is $32800$.

\subsection{EtherNet Galaxy: a Software for Blockchain Analysis}
The complex networks analysis described above has been performed using a new software called EtherNet Galaxy. A complete description of EtherNet Galaxy is beyond the scope of this paper but in the following a brief description of the main EtherNet Galaxy features is reported. Firstly, EtherNet Galaxy retrieves the Ethereum blockchain data using the APIs provided by Infura (https://infura.io/), a service that delivers RPC access to the Ethereum network. Thanks to this service, EtherNet Galaxy is able to retrieve the information about the blocks (e.g.~block number, size, list of transactions, etc.). Secondly, the retrieved blocks are analyzed using the web3-eth package (https://github.com/ethereum/web3.js/). In this stage, the goal is to extract all the transaction encoded in each block and to represent them as a network using the Pajek data format. Finally, the network analysis on the previously generated graphs is performed by EtherNet Galaxy relying on the Python NetworkX software library~\cite{networkx}. The EtherNet Galaxy network analysis software is currently under development, a prototype version has been used for the analysis reported in the following of this paper. At a later stage, EtherNet Galaxy will be made freely available on the research group homepage (https://site.unibo.it/anansi/).

\section{Results}\label{sec:results}

\begin{table*}[h]
\caption{General Metrics of Considered Networks.}
\begin{center}
	\begin{tabular}{ | c | c |  c | c | c | c | c |}
	\hline
  \textbf{\# Blocks} & \textbf{\# nodes} & \textbf{\# edges} & \textbf{avg clus coeff}  &
                     \textbf{\# components} &
                     \textbf{\# nodes largest comp} & \textbf{\# edges largest comp}\\ \hline
  1     &   55  &   40  &   0   &   15  &   19  &   18\\ 
  10    &   846 &   648 &   0   &   199 &   139 &   139\\
  $10^2$   &   7507    &  7184 &  0.001   &   729 &   4262    &   4641 \\
  $10^3$  &   47469   &  51357    &   0.006   &   2848    &   37201   &   43682\\
  $10^4$ &   284630  &   347679  &   0.014   &   10770   &   239114  &   303248\\
  $10^5$    &   1467960 &   2144095 &   0.036   &   40276   &   1321468 &   1994428\\
	\hline
  \end{tabular}
\end{center}
\label{table:general}
\end{table*}

Table~\ref{table:general} shows some main metrics related to six networks obtained by considering the set of transactions contained in different numbers of blocks, i.e.~$1$, $10$, $100$, $1000$, $10000$, $100000$. A the time of writing, the Ethereum blockchain explorers, e.g.~https://bitinfocharts.com/ethereum/, report that the average time between blocks is $14.1$ sec, while the average number of blocks per hour is $254$. Thus, we can roughly state that the considered networks are related to numbers of transactions ranging from few seconds up to $16$ days.

As we expected, if we consider transactions that are contained in a single block, we obtain a very simple network, with few nodes and few edges. The number of nodes is higher than the number of edges; we might thus expect that there are transactions with multiple recipients. Due to the (essentially) random nature of the choice of transactions inserted in a block, we can imagine that the transactions involve different nodes. Thus, the resulting network is very sparse. Indeed, there are no triangles in the network (the clustering coefficient is zero). The amount of network components\footnote{A component is a sub-graph of the main net, in which any two nodes are connected to each other by paths, i.e.~the component is composed of nodes connected through links.} is relatively high, with respect to the number of nodes in the network.

\begin{figure}
\centering
  \begin{subfigure}
    \centering
    \includegraphics[width=4cm]{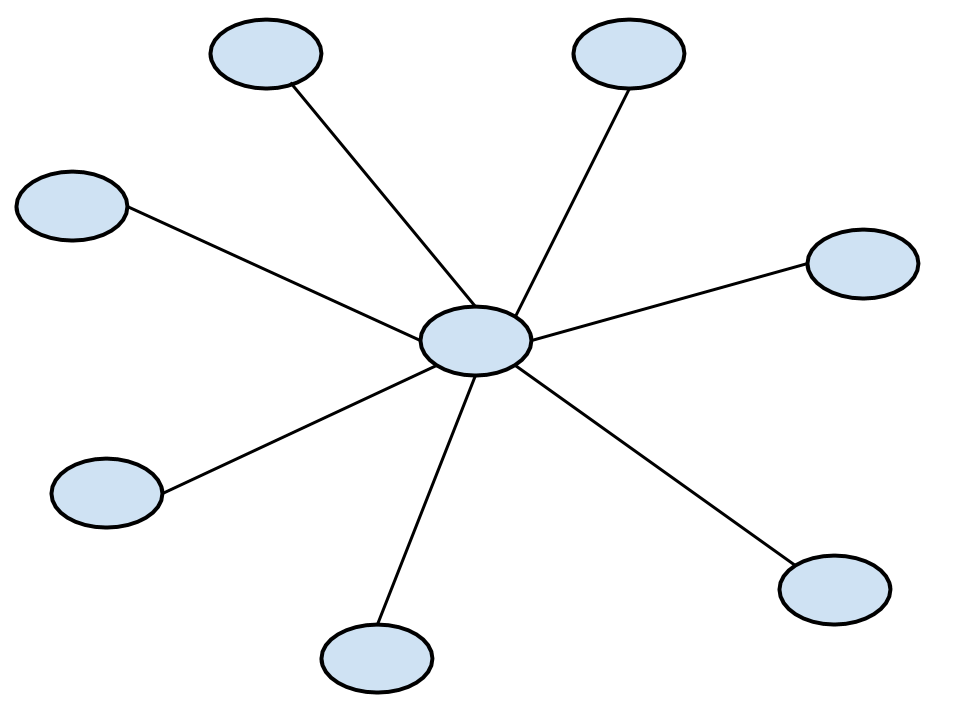}
    \caption{star}
    \label{fig:star}
  \end{subfigure}
  \begin{subfigure}
    \centering\includegraphics[width=4cm]{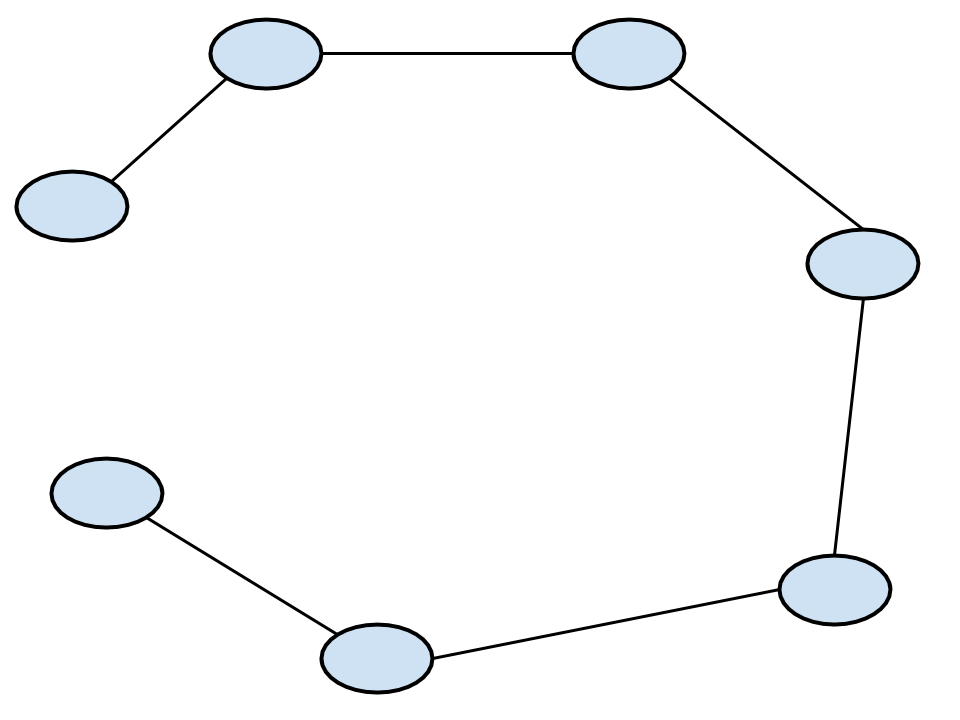}
    \caption{chain}
    \label{fig:chain}
  \end{subfigure}
  \caption{Different interaction patterns among Ethereum addresses, depicted as a network.}
\end{figure}

Let's consider the main component of the $1$-block network. It is composed of $19$ nodes and $18$ links (as reported in the table). We already stated that this network corresponds to a bunch of transactions included in a single block i.e.~probably generated in a small time interval. We might have two options here. The first one is that the component has a star structure, meaning that a specific address had an interaction with a set of nodes as shown in Figure~\ref{fig:star}. Indeed, star structures appear frequently also in the wider network reported in Figure~\ref{fig:snap}. The second option is that the interactions correspond to a chain of different transactions (Figure~\ref{fig:chain}). We think that this alternative is more unlikely, since it would mean that during the block generation, the miner selected a bunch of transactions, generated in a given time interval, involving a chain of accounts.

As concerns isolated pairs of connected nodes, or small sized components, these might represent few (test) transactions among different accounts. Alternatively, they might represent the deployment of some (prototype) smart contracts, that is indeed realized through the triggering of a specific transaction in the blockchain, plus some possible (test) interactions with the smart contract. Indeed, at the time of writing, this is a quite common use of the Ethereum blockchain.

When we consider wider networks, built based on higher amounts of blocks, values of the considered metrics increase considerably. Still, all of them show a low average clustering coefficient. All networks show a high number of components, with respect to the amount of nodes. However, when we increase the number of blocks the main component embodies a high majority of network nodes (e.g.~$\sim90\%$ for the biggest net).

\begin{figure*}
\centering
  \includegraphics[width=.39\linewidth]{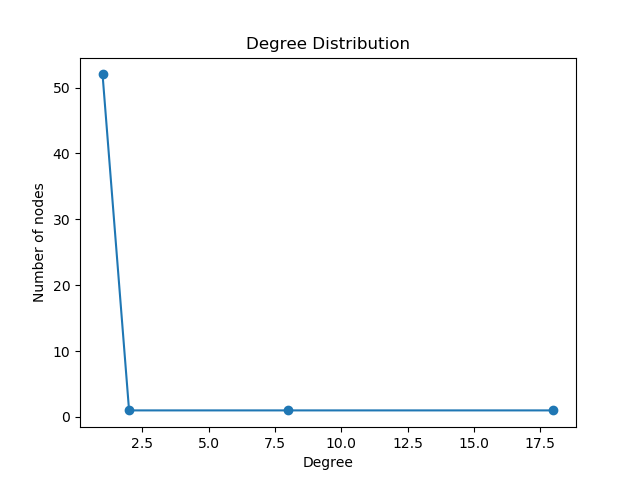}
  \includegraphics[width=.39\linewidth]{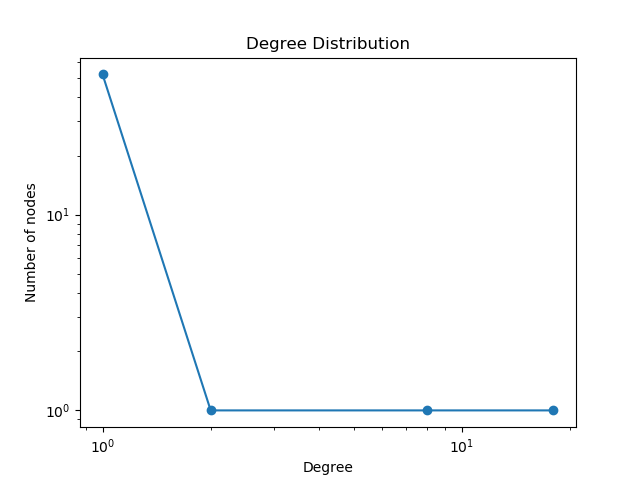}
  \caption{Degree distribution in linear and log scales -- 1 block.}
  \label{fig:res_1b}
\end{figure*}

\begin{figure*}
\centering
  \includegraphics[width=.39\linewidth]{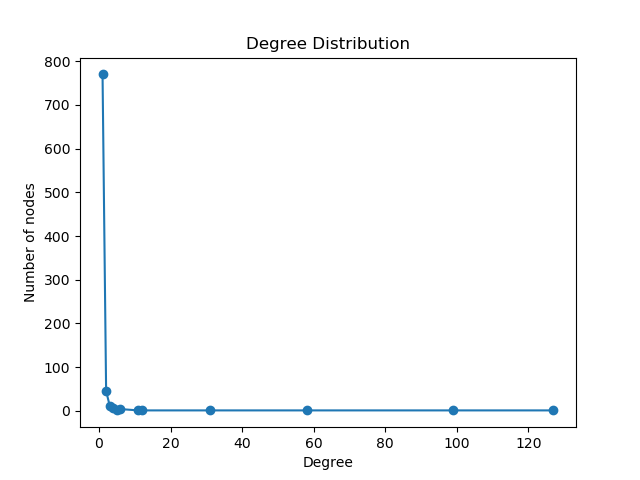}
  \includegraphics[width=.39\linewidth]{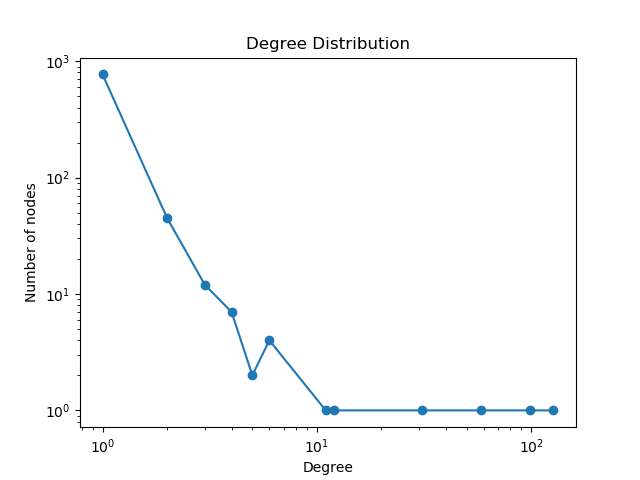}
  \caption{Degree distribution in linear and log scales -- 10 blocks.}
  \label{fig:res_10b}
\end{figure*}

\begin{figure*}
\centering
  \includegraphics[width=.39\linewidth]{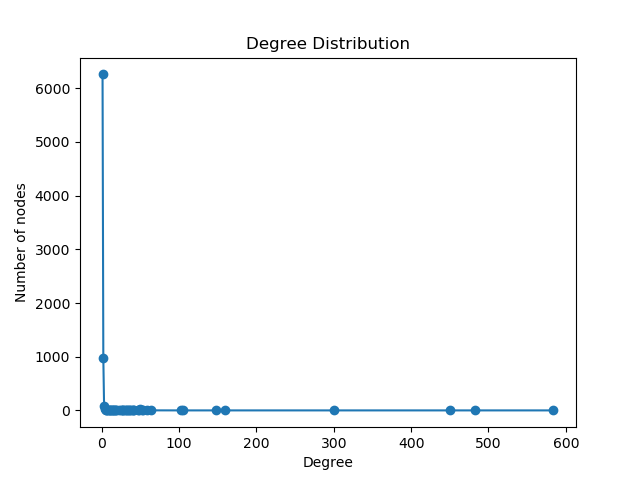}
  \includegraphics[width=.39\linewidth]{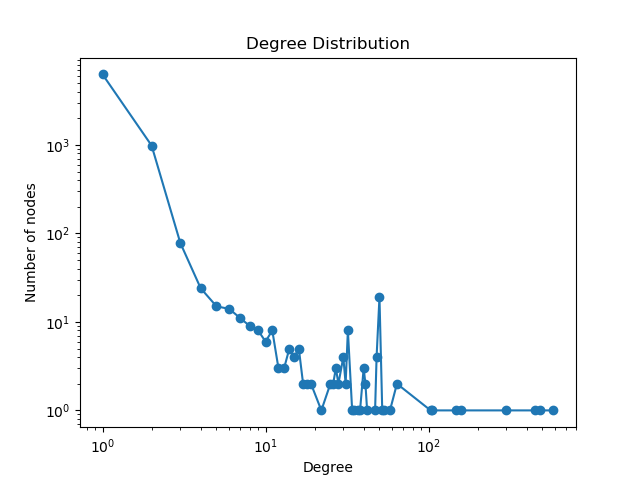}
  \caption{Degree distribution in linear and log scales -- 100 blocks.}
  \label{fig:res_100b}
\end{figure*}

\begin{figure*}
\centering
  \includegraphics[width=.39\linewidth]{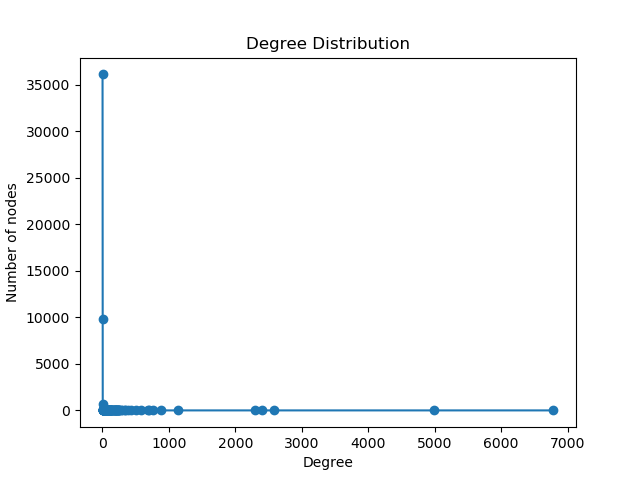}
  \includegraphics[width=.39\linewidth]{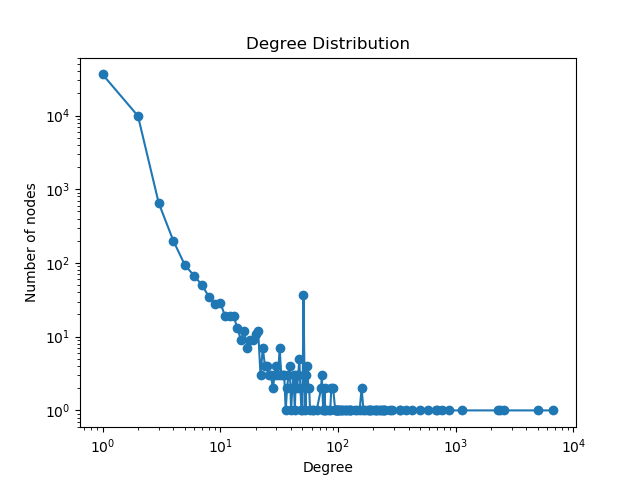}
  \caption{Degree distribution in linear and log scales -- 1000 blocks.}
  \label{fig:res_1000b}
\end{figure*}

\begin{figure*}
\centering
  \includegraphics[width=.39\linewidth]{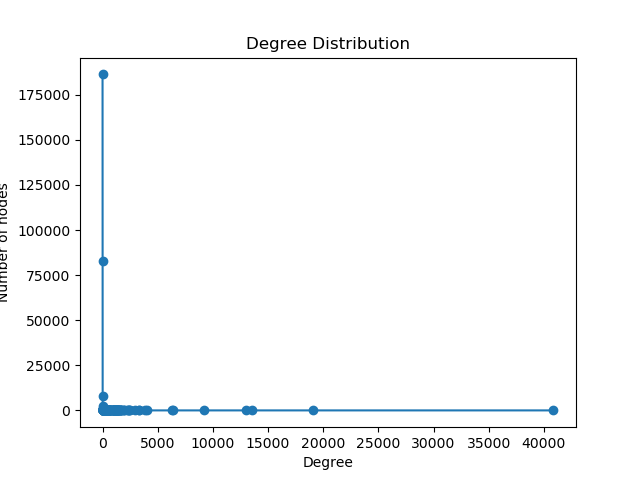}
  \includegraphics[width=.39\linewidth]{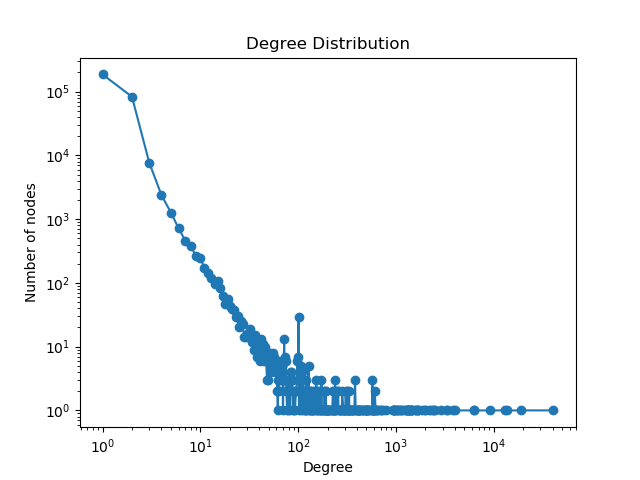}
  \caption{Degree distribution in linear and log scales -- 10000 blocks.}
  \label{fig:res_10000b}
\end{figure*}

\begin{figure*}
\centering
  \includegraphics[width=.39\linewidth]{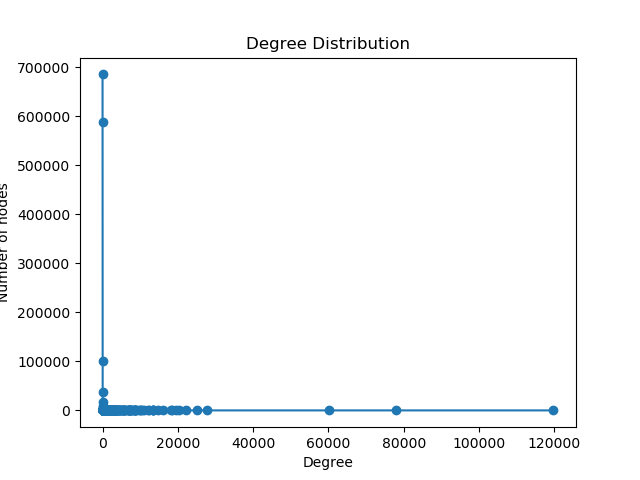}
  \includegraphics[width=.39\linewidth]{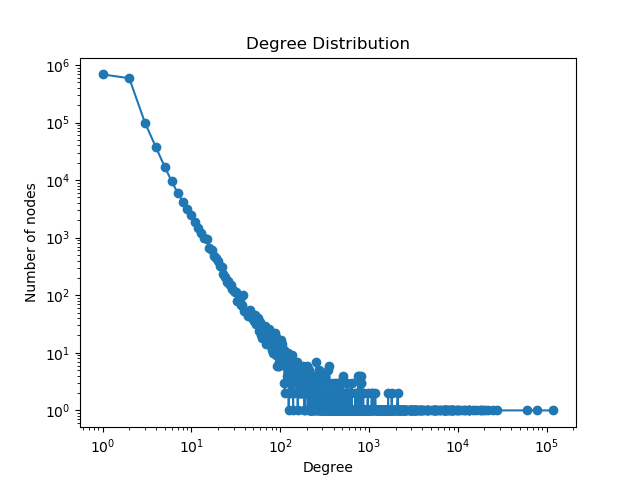}
  \caption{Degree distribution in linear and log scales -- 100000 blocks.}
  \label{fig:res_100000b}
\end{figure*}

\subsection{Degree Distribution}

Figures~\ref{fig:res_1b}--\ref{fig:res_100000b} show the degree distributions of different networks obtained by analyzing an increasing number of $10^n$ blocks, with $n=0,\dots,5$; i.e.~from 1 up to $10^5$ blocks, respectively. In each figure, we report the degree distribution using a linear scale (left chart) as well as in a log-log scale (right chart). The log-log chart is interesting, since it easily allows to understand if, for instance, the degree distribution follows a power law function (in this case the plot of the degree distribution should appear as a straight line), rather than an heavy tailed distribution, etc.

With just one block, the obtained network is quite simple. There is a limited number of nodes that created transactions (included in the block). Moreover, it is quite unlikely that an account is involved in more than one transaction per block. The degree distribution in Figure~\ref{fig:res_1b} confirms this.

When we increase the order of magnitude of considered blocks, then things start changing. Still, the high majority of nodes performed a single transaction (i.e.~they have a degree equal to $1$). However, the percentage of nodes with higher degrees (i.e.~higher amounts of transactions with different nodes) increases. If we look at the chart in log-log scale, we can see an almost linear decrease on the distribution of the degrees, with a long tail, suggesting that those degrees follow a power law function.

It is important to mention that a common practice in cryptocurrencies, especially in Bitcoin, is to create a fresh address for each payment a user receives. This in order to decouple the recipient of different transactions and increase the level of anonymity. Indeed, the wallet of a cryptocurrency is enabled to manage different user addresses/accounts, and a new address makes it more difficult to trace the cryptocurrency trail. Actually, most online wallets automatically create a new address each time a user is involved as the output of a transaction. Moreover, creating a new address for each transaction is a fool-proof way of ensuring that someone has paid the user, because the user gave that address to only that person and no one else.

\subsection{Small World Phenomenon}

To assess the small world property of the considered networks, we compare the clustering coefficient and the average path length of their main component with that of the equivalent random graph (generated by taking the same amount of nodes, with the same amount of edges randomly distributed among these nodes). Thank to these values, we are able to compute the $\sigma$ value (Equation~\ref{eq:sigma}). All these measures are reported in Table~\ref{tab:sw}.

\begin{table*}[ht]
\caption{Small worlds: Ethereum networks vs. related random graphs. (Results on main components.)}
\begin{center}
	\begin{tabular}{ | c | c |  c | c | c | c | c | c |}
	\hline
  \textbf{\# Blocks} & \textbf{\# nodes} & \textbf{\# edges} & \textbf{cc}  & \textbf{L} &
                     \textbf{cc RG} & \textbf{L RG} & $\sigma$ \\ \hline
  1     &   19  &  18  &  0  &  1.89  &  0.05 &  2.94  & 0\\
  10    &   139 &  139 &  0  &  2.26  &  0.01 &  4.94  & 0\\
  $10^2$   &  4262  &   4641  &  0  &  7.45   &   0  &  8.36  &  2.07\\
  $10^3$  &   37201 &   43682 &  0.003  &  5.66   &  $3\times$ ${10^-5}$   &  10.52 &  225.85\\
  $2\times10^3$  &    67911 &   81580 &  0.005  &  5.24   &  $1\times$ ${10^-5}$   &  11.13 &  602.16\\
  $3\times10^3$  &    88184  &   106842 &  0.01  &  5.28   &  $1.37\times$ ${10^-5}$   &  11.39 &  1520.28\\
  $4\times10^3$  &    108871  &   108871 &  0.01  &  5.29   &  $1.12\times$ ${10^-5}$   &  11.60 &  1770.93\\
  $5\times10^3$  &    133982  &   164406 &  0.01  &  5.35   &  $9.15\times$ ${10^-6}$   &  11.80 &  2272.96\\ 
  $10^4$ &   239114 & 303248    &   0.01    &   5.28    &   $5.30\times$ ${10^-6}$   &  12.38 &  5891.43\\
	\hline
  \end{tabular}
\end{center}
\label{tab:sw}
\end{table*}

As already mentioned, with a low amount of blocks, a limited set of transactions is considered. Thus, we have very simple networks with small main components. These components have a null clustering coefficient ("cc" column in Table \ref{tab:sw}). It is clear that these networks are not small worlds. Also when we increase the amount of blocks, the clustering coefficients of the obtained networks is almost zero. This allows to conclude that even bigger networks are not small worlds. As concerns the $10^3$-blocks network, the clustering coefficient of the random graph, obtained with the same amount of edges of the original network, is so small the the final $\sigma$ value is really high. However, we claim that this is just a numerical outcome and it does not justify stating that this network is a small world.

The fact that the considered networks cannot be associated to small worlds is confirmed by looking at their related diameters, which are reported in Table~\ref{tab:diam}. It is worth noticing that the diameter is calculated on the main component only. When we increase the network size, the diameter increases as well. The increment of the diameter is particularly evident when passing from the $10$-blocks network to $10^2$-blocks one. In this last case, we notice a network diameter of $23$ that is quite far from the  ``six degrees of separation'', that is usually associated to small worlds~\cite{Ferretti20131631}.

\begin{table}[h]
\caption{Networks diameter.}
\begin{center}
	\begin{tabular}{ | c | c | c |}
	\hline
  \textbf{\# Blocks} & \textbf{\# nodes (main comp)} & \textbf{diameter} \\ \hline
  1     &   19  &  2\\
  10    &   139 &  4\\
  $10^2$   &  4262  &   23\\
  $10^3$  &   37201 &   27\\
  $2\times10^3$  &    67911 &   22\\
  $3\times10^3$  &    88184 &   24\\
  $4\times10^3$  &    108871 &   34\\
  $5\times10^3$  &    133982 &   58\\
  $10^4$ &   239114 &   90\\
	\hline
  \end{tabular}
\end{center}
\label{tab:diam}
\end{table}

\subsection{Metrics at Different Snapshots}

As a further analysis, we tried to understand if the Ethereum network, obtained using a set of blocks, changes over time, i.e.~we tried to understand if the evolution of the Ethereum blockchain has some effects on the related networks. To answer this question, we took different slices of the blockchain, by considering different snapshots starting at block numbers: 1000000, 2000000, 3000000, 4000000, 5000000, 6000000 and 7000000.
The size of each of these blockchain slices are $1000$ blocks (i.e.~approximately 4 hours).

\begin{figure*}
\centering
  \includegraphics[width=.39\linewidth]{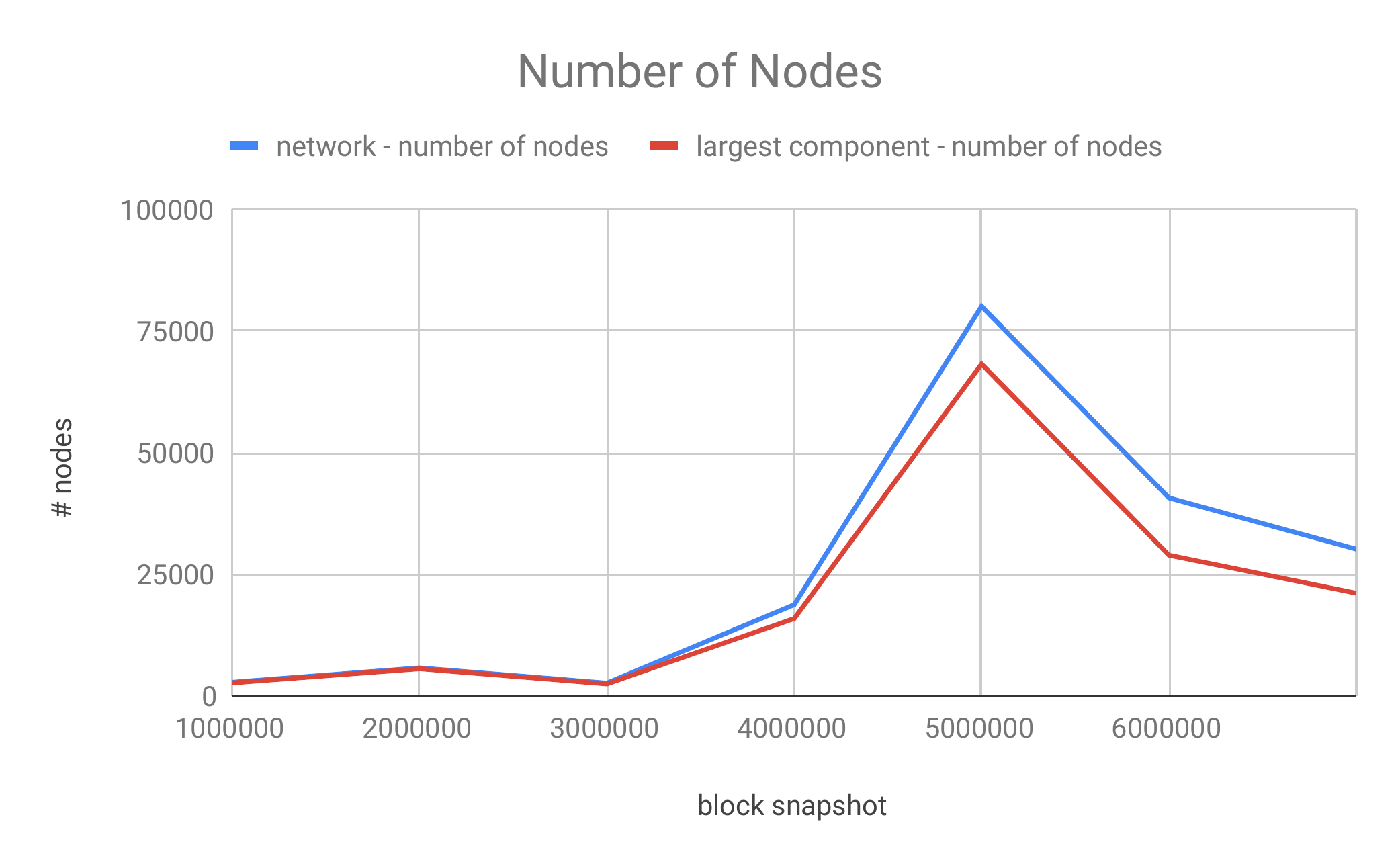}
  \includegraphics[width=.39\linewidth]{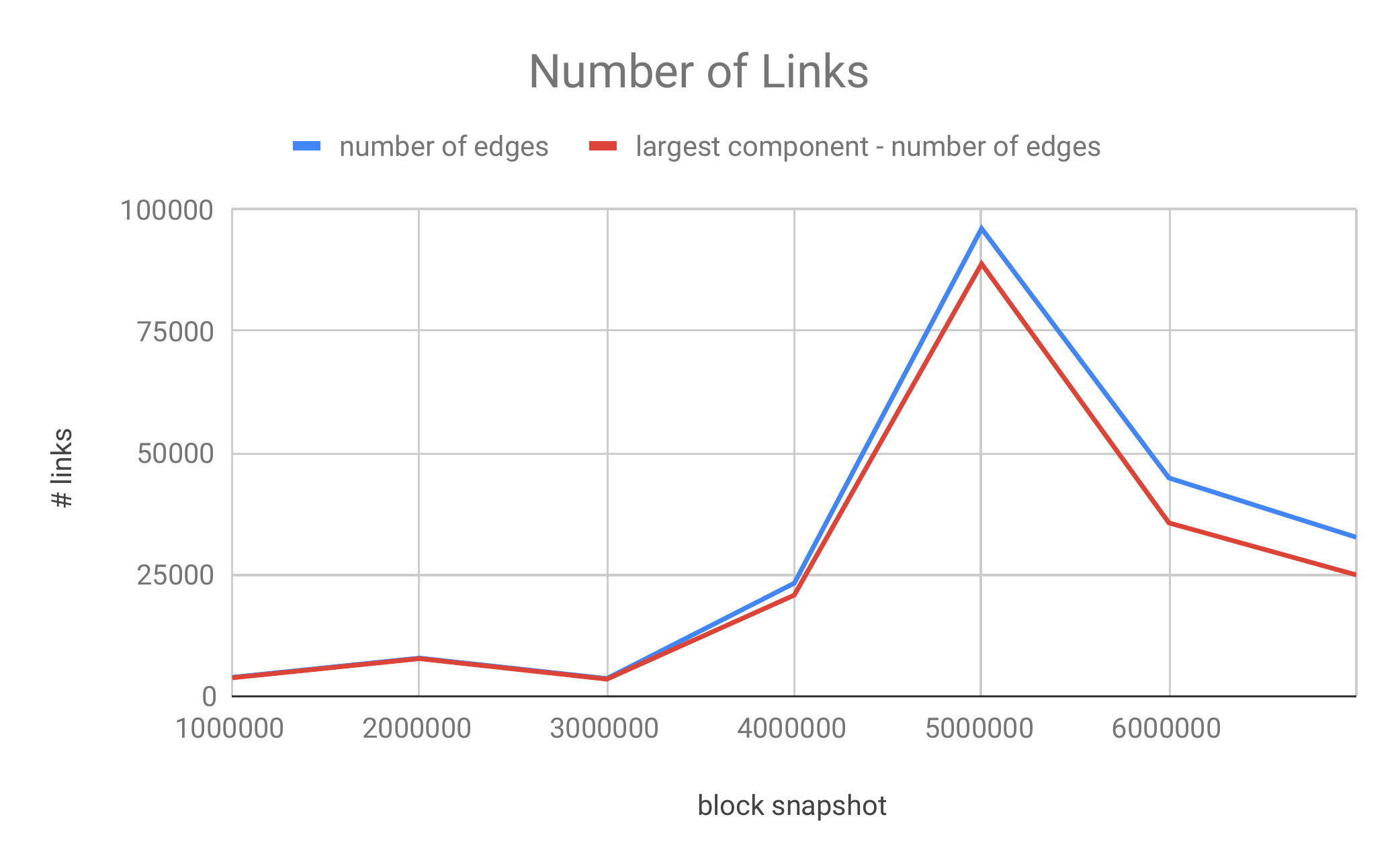}
  \includegraphics[width=.39\linewidth]{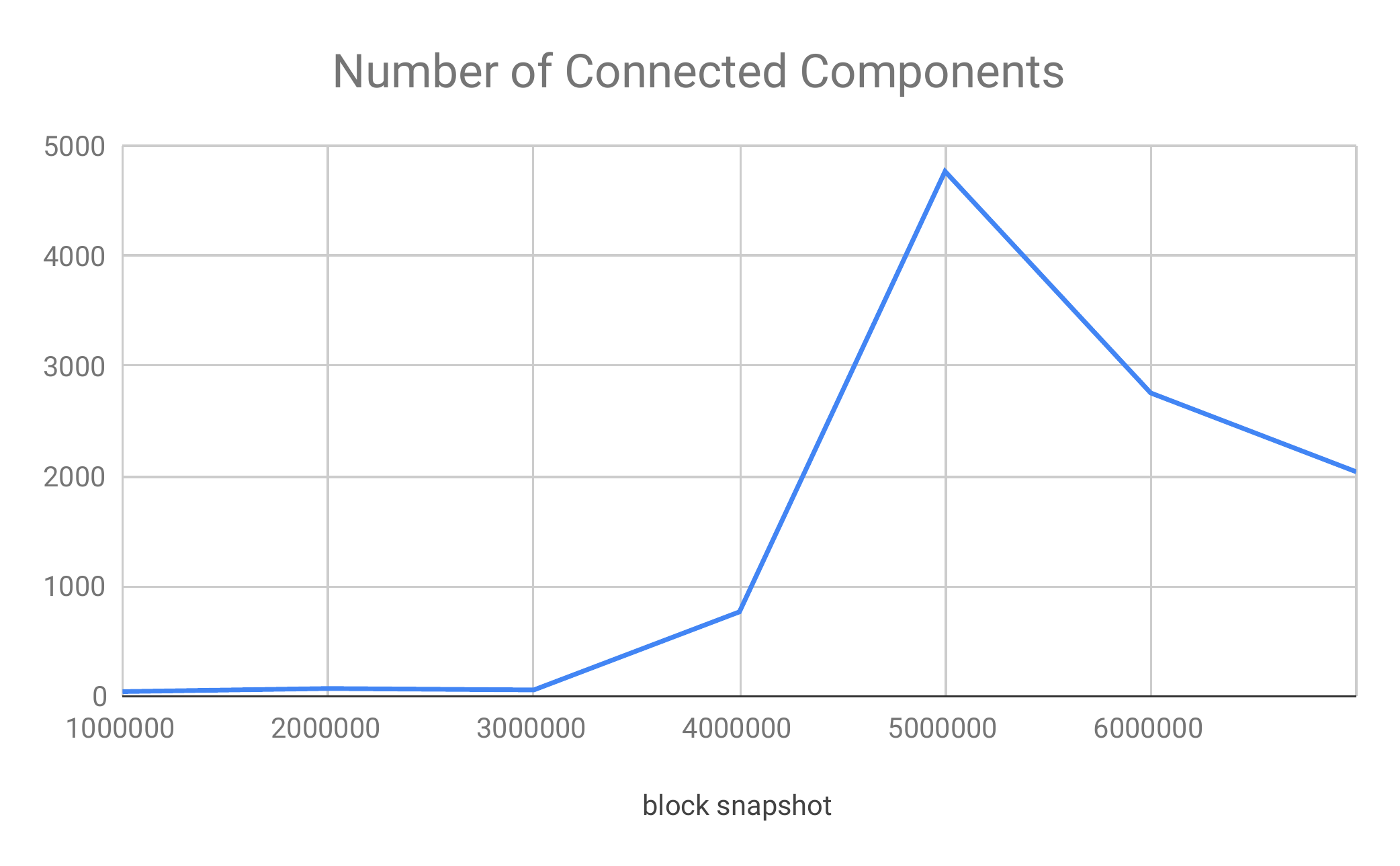}
  \includegraphics[width=.39\linewidth]{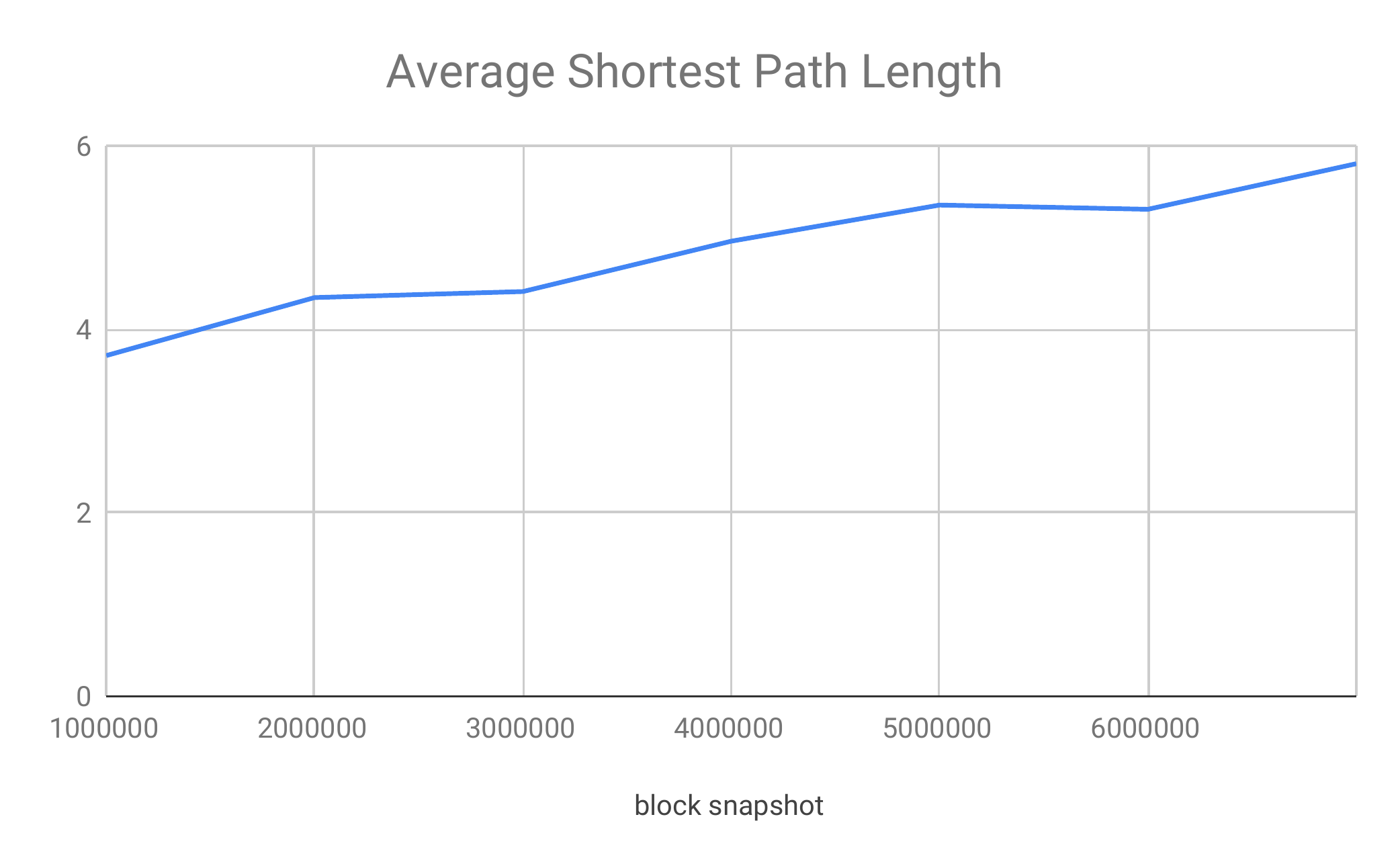}
  \caption{Network statistics at different block snapshots -- snapshot size: 1000 blocks.}
  \label{fig:1000_snapshost}
\end{figure*}

\begin{figure}
\centering
  \includegraphics[width=.8\linewidth]{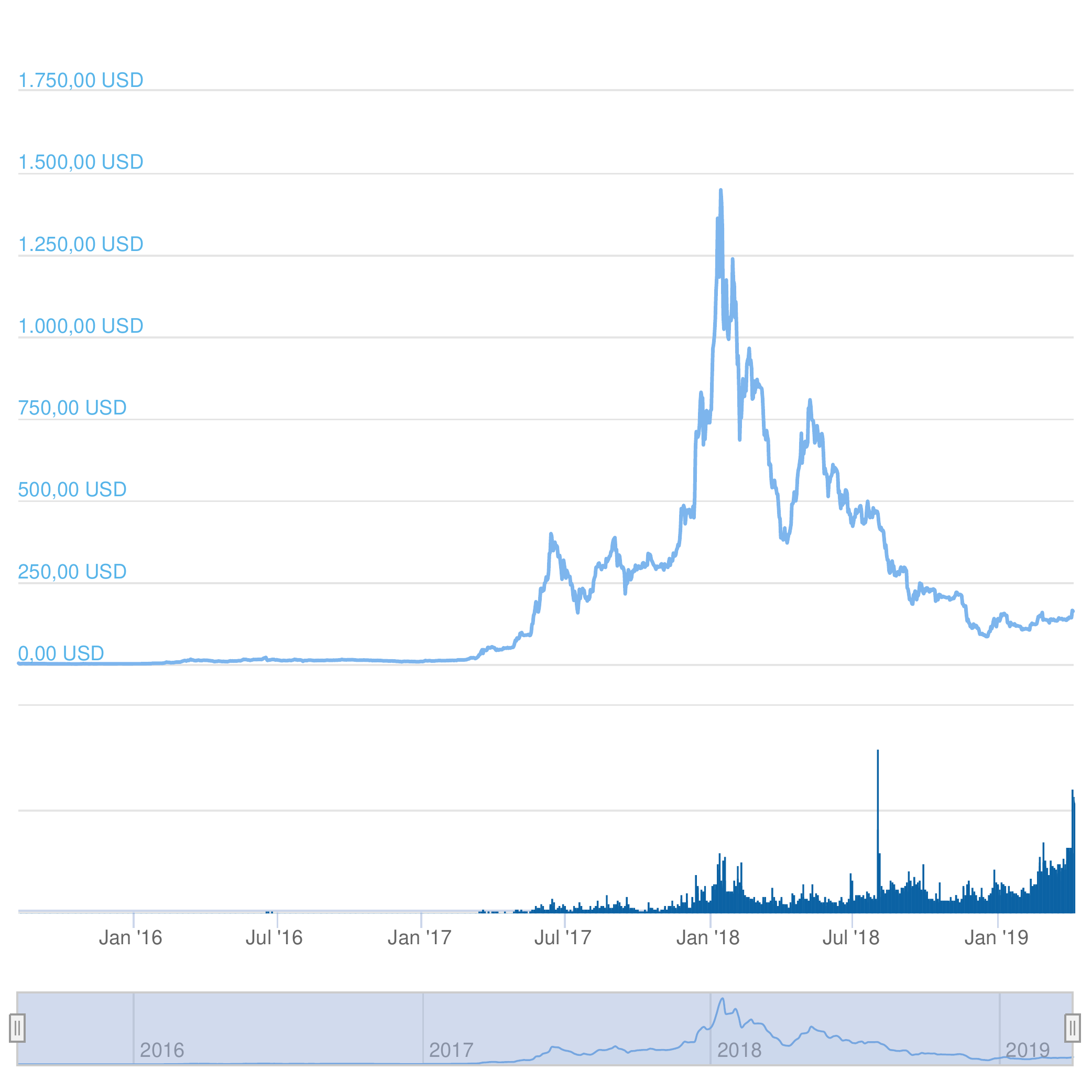}
  \caption{Eth-USD exchange value -- source: https://www.coingecko.com.}
  \label{fig:exchange}
\end{figure}

Figure~\ref{fig:1000_snapshost} shows different network metrics, when we consider networks obtained from different points of the blockchain. These metrics are the amount of nodes (we report both the total amount of nodes in the net and the number of nodes in the main component), the amount of links (both in the whole network and in the main component), the number of components of the network and the average shortest path length (in the main component). In general, all these charts demonstrate how the network has grown in time, in terms of nodes and interactions. We can notice a spike in the snapshot took at $5000000$. It seems that in that slice, a notable amount of interactions occurred. This might be explained by the fact that the considered blocks, in the snapshot identified as $5000000$, have been generated in date Jan-30-2018. Indeed, if we look at the exchange rate Ether - US Dollar, shown in Figure~\ref{fig:exchange}, in that time period we will notice a spike in the Eth value. It is reasonable to assume that the higher the value of the crypto-currency linked to the blockchain, the higher the activity in the blockchain.

\begin{figure*}
\centering
  \includegraphics[width=.39\linewidth]{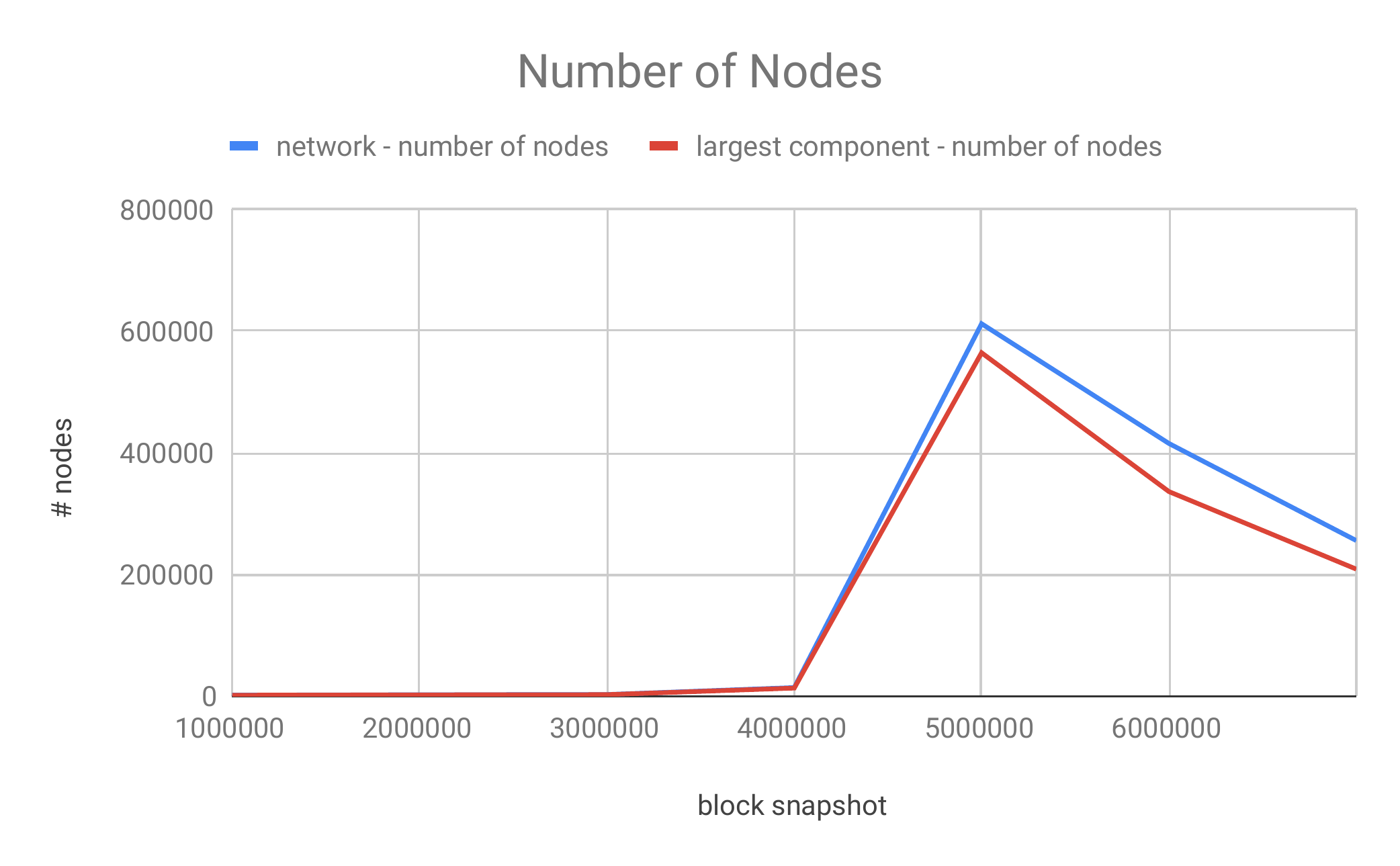}
  \includegraphics[width=.39\linewidth]{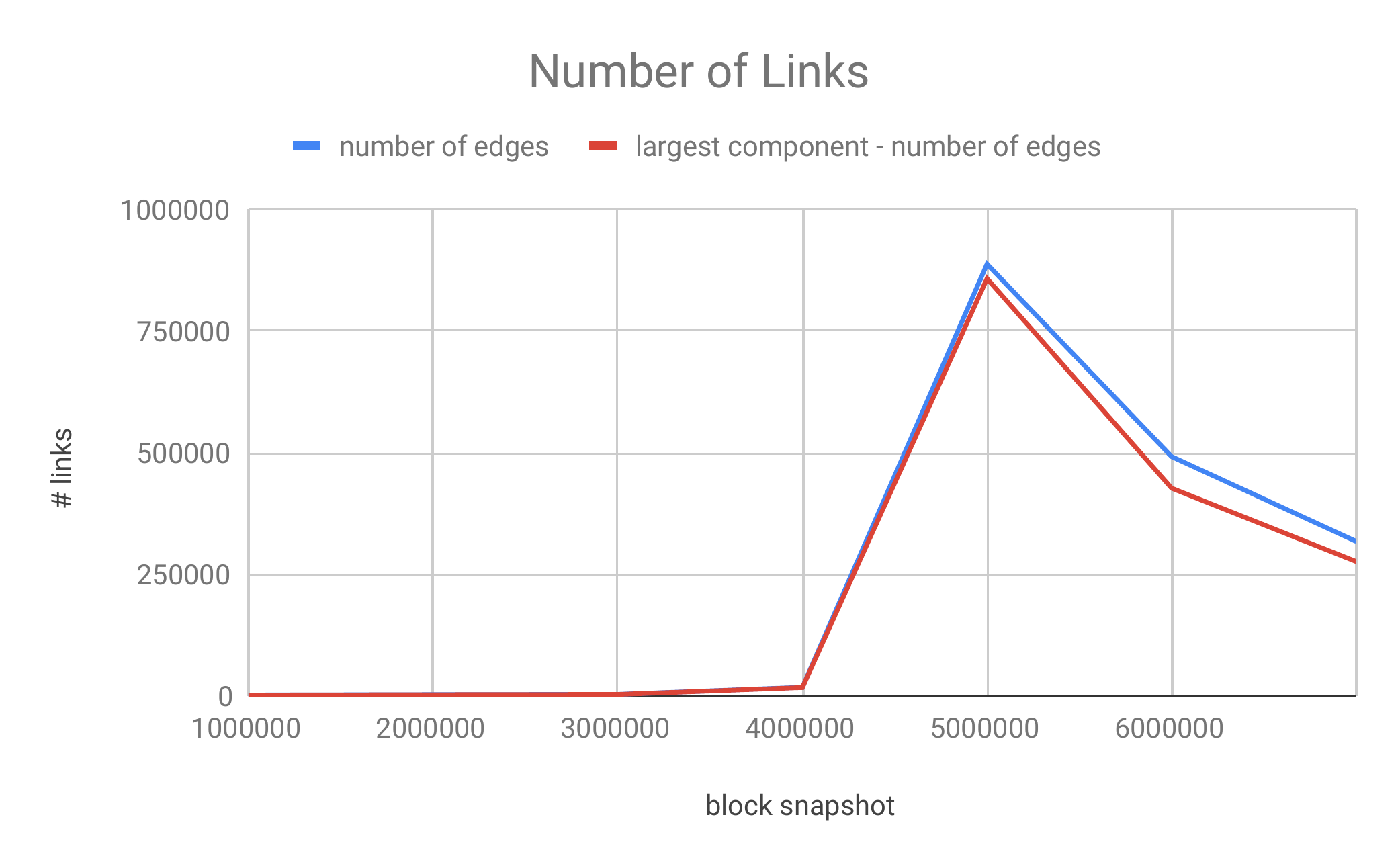}
  \includegraphics[width=.39\linewidth]{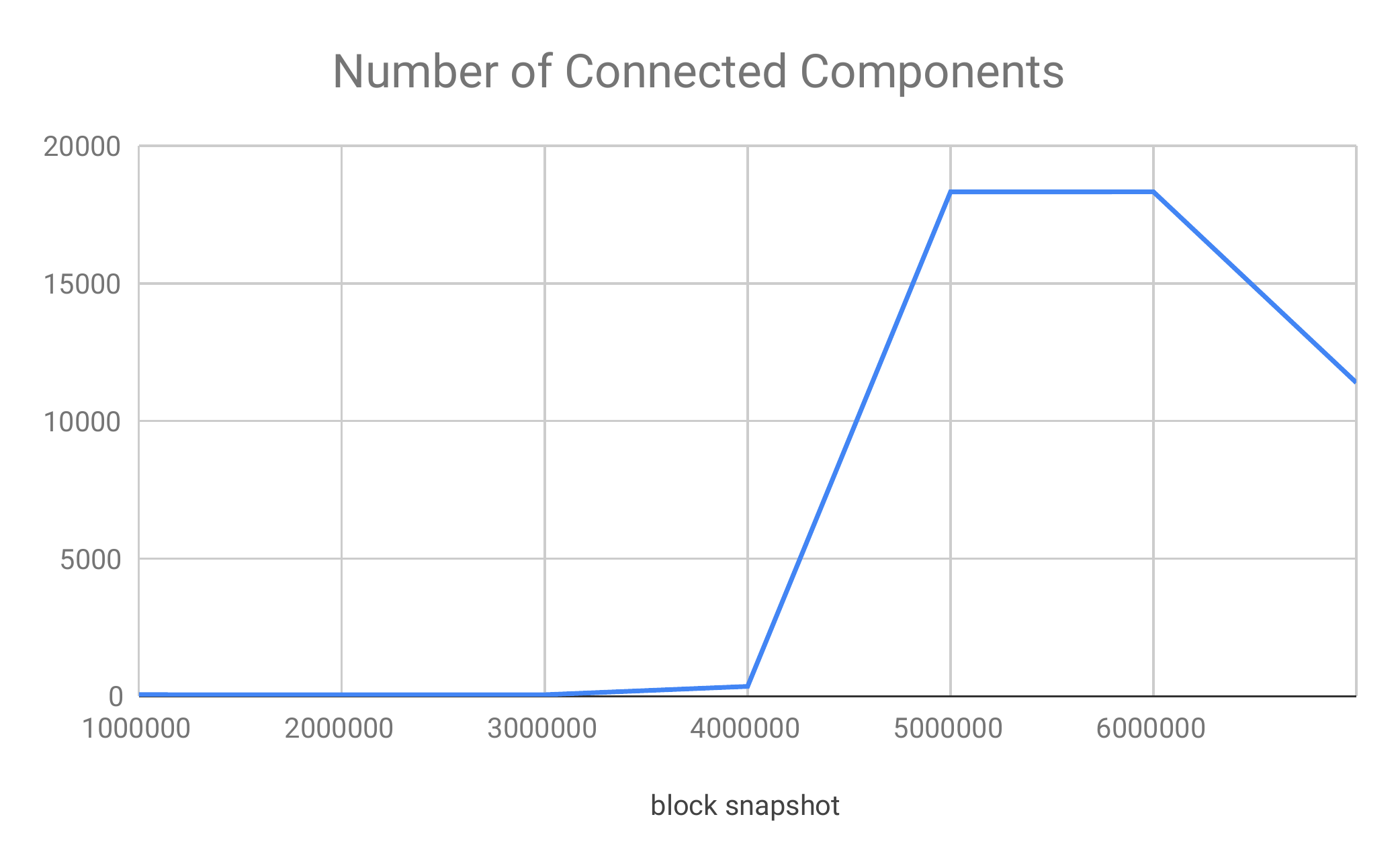}
  \includegraphics[width=.39\linewidth]{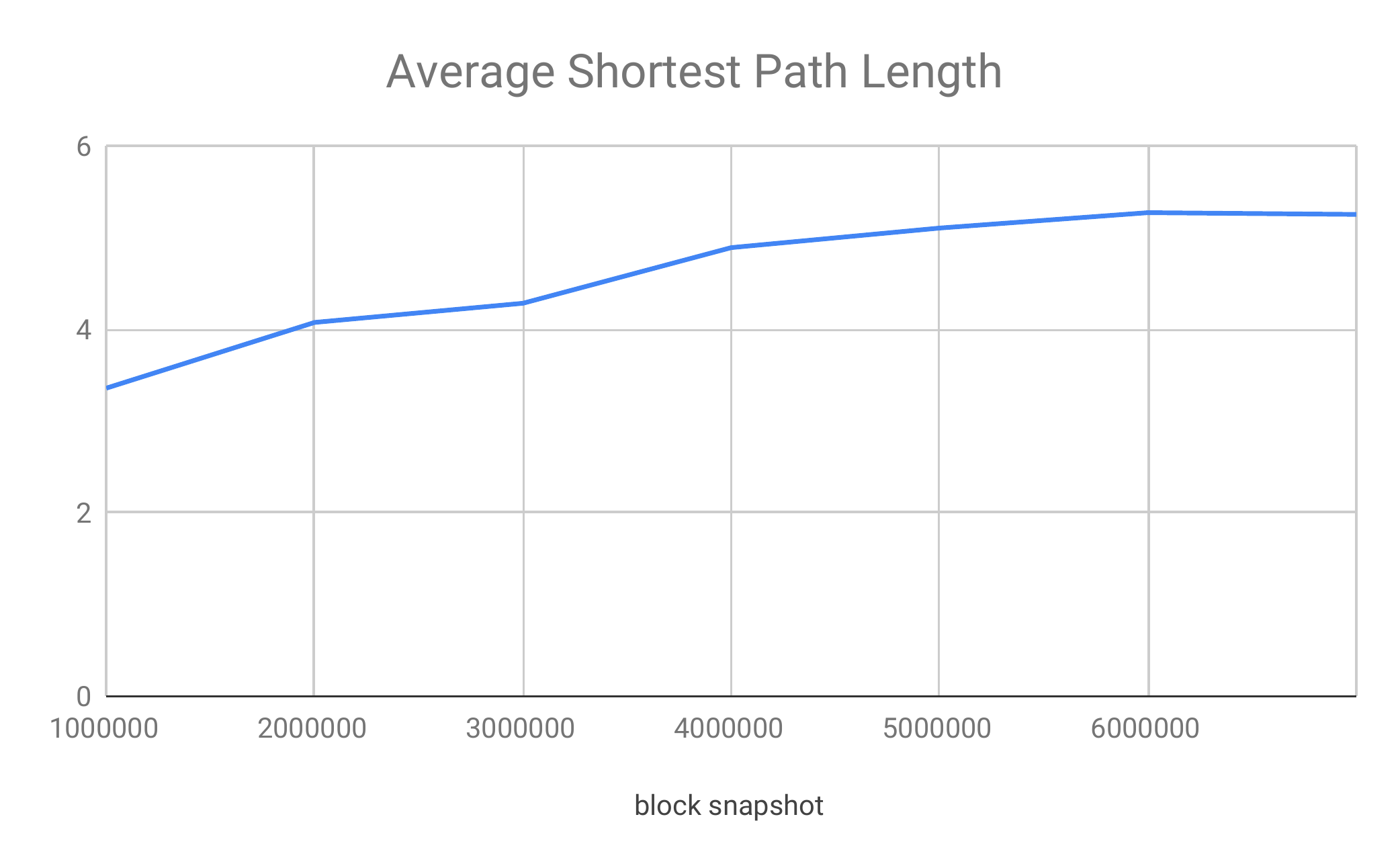}
  \caption{Network statistics at different block snapshots -- snapshot size: 10000 blocks.}
  \label{fig:10000_snapshost}
\end{figure*}

Figure \ref{fig:10000_snapshost} shows results similar to those of Figure \ref{fig:1000_snapshost}, but when the blocks intervals' was of $10000$ blocks, i.e.~approximately 40 hours. Values are different, but the trend is analogous to results related to $1000$ blocks' slices.

It is worth mentioning that we measured other metrics as well, such as the average clustering coefficient. In this case, we did not noticed a specific trend. All networks have a very low average clustering coefficient, similar to the values show in previous subsections.

\subsection{Miners Distribution}

The analysis of the blockchain allows retrieving diverse typologies of information, related to the generation of blocks. Besides the number of accounts that interact in the blockchain, another interesting metrics is concerned to miners that generate blocks. In particular, it might be interesting to understand if the distribution of miners is truly spread out, or rather if a small niche of miners have the control of the blockchain.

\begin{figure*}
  \centering
  \includegraphics[width=.65\linewidth]{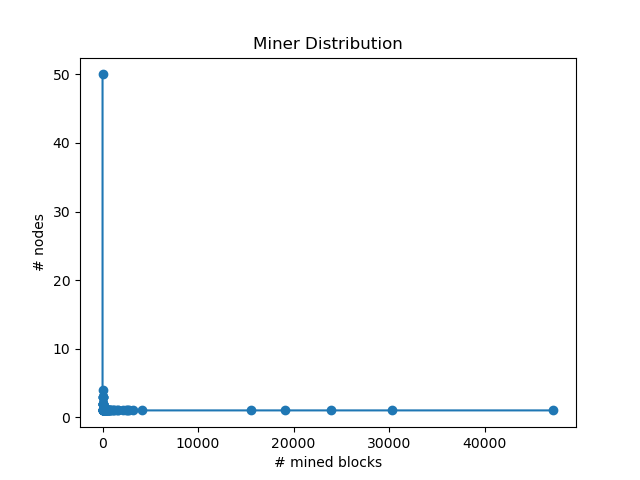}
  \caption{Distribution of of nodes that mined a certain number of blocks.}
  \label{fig:miners}
\end{figure*}

Figure \ref{fig:miners} shows the distribution of nodes that mined a certain number of blocks. On the x-axis there is the amount of blocks that have been mined by the same miner. On the y-axis, we have the amount of nodes that mined that number of blocks. In this case, we considered a subset of $\sim180000$ blocks, starting backwards from the last available block in the blockchain at the time of the conducted experiments, i.e.~the most recent block was the same of results reported at on the first part of this section.

It is possible to observe that, while as expected the majority of miners were able to mine just one block in the considered time interval, there are however certain nodes that have mined a huge number of blocks. In particular, it seems that six nodes mined over $10000$ blocks; one node mined $47193$ blocks. These nodes are probably mining pools, i.e., a set of nodes who share their processing power, with the aim to split the reward equally, according to the amount of work they contributed to the probability of finding a block. This obtained result is in line with the statistics offered by blockchain explorer web sites, e.g., https://www.etherchain.org/.

\section{Conclusions}\label{sec:conc}

In this paper, we provided an analysis of the complex network representing the Ethereum blockchain transactions. We varied the amount of blocks considered to build the network. This corresponds to a different temporal range, thus to a different amount of transactions and a different network size. As expected, the wider the network the more likely the presence of hubs in the network, meaning that there are some nodes that are more active in the blockchain. We also considered different temporal intervals, by taking subsets of subsequent blocks back in the chain. This allows to understand how the use of the blockchain changes in time.

Even if Ethereum accounts are anonymous, and hence it is not possible to directly map an external account to a given user, we might assume that hub nodes correspond to well known accounts, that might represent popular smart contracts or external accounts that allow exchanging Ether. Examples are those services that require a real identity to transact, such as  online wallet services, currency exchange services, merchants. In public blockchains, such as Ethereum, account anonymity is obtained by employing pseudonymity to represent an account, together with the unlinkability among different interactions of the same user with the system. If an account is a hub node in the network, and it is not a popular smart contract, this means that the related real world entity often employs the same account to transact. In this case, it might be easier to de-anonymize such an account~\cite{Meiklejohn:2013}.

\small{
\bibliographystyle{abbrv}
\bibliography{paper}  
}

\end{document}